\documentclass[11pt]{article}
\usepackage[margin=1truein]{geometry}

\usepackage{setspace}
\usepackage{lscape}
\usepackage{pdflscape}

\usepackage{amsmath,amssymb,amsthm} 

\usepackage{wrapfig} 

\usepackage{mathrsfs} 
\usepackage{url} 
\usepackage{latexsym} 
\usepackage{bm} 

\usepackage{mathtools}

\usepackage{booktabs,multirow}
\usepackage{array}
\newcolumntype{P}[1]{>{\centering\arraybackslash}p{#1}}
\usepackage{diagbox}

\usepackage{algorithm}

\usepackage[sort&compress,numbers]{natbib}

\usepackage{thmtools,thm-restate}

\usepackage{hyperref}
\usepackage{cleveref}
\expandafter\def\csname ver@etex.sty\endcsname{3000/12/31} 
\usepackage{autonum} 
\makeatletter
\autonum@generatePatchedReferenceCSL{Cref} 
\makeatother

\usepackage{tikz,tikz-3dplot}
\usetikzlibrary{positioning,shapes,shapes.callouts,shapes.multipart,arrows.meta,fit,calc}
\usepackage{pgfplots}
\pgfplotsset{compat=newest}
\usepackage{pxpgfmark} 
\usetikzlibrary{bayesnet} 

\usepackage{multicol}
\usepackage{wasysym} 
\usepackage{adjustbox}
\usepackage{xparse} 
\usepackage{pbox} 
\usepackage{authblk} 
\usepackage{xspace} 

\usepackage[subrefformat=parens]{subcaption}
\usepackage[inline]{enumitem} 
\hypersetup{
  colorlinks,
  linkcolor={green!35!black},
  citecolor={green!35!black},
  urlcolor={blue!50!black}
}

\DeclareMathOperator*{\argmax}{argmax}
\DeclareMathOperator*{\argmin}{argmin}

\DeclareMathOperator{\EE}{\mathbb{E}}

\DeclarePairedDelimiterX\inner[2]{\langle}{\rangle}{{#1},{#2}}

\DeclarePairedDelimiter\set{\{}{\}}
\DeclarePairedDelimiter\prn{(}{)}
\DeclarePairedDelimiter\bra{[}{]}
\DeclarePairedDelimiterX\Set[2]{\{}{\}}{\mspace{2mu}{#1}\;\delimsize|\;{#2}\mspace{2mu}}
\DeclarePairedDelimiterX\Prn[2]{(}{)}{\mspace{2mu}{#1}\;\delimsize|\;{#2}\mspace{2mu}}
\DeclarePairedDelimiterX\Bra[2]{[}{]}{\mspace{2mu}{#1}\;\delimsize|\;{#2}\mspace{2mu}}

\newcommand{\Z}{\mathbb Z}
\newcommand{\R}{\mathbb R}

\newcommand{\st}{\mathrm{s.t.}}

\renewcommand{\epsilon}{\varepsilon}

\NewDocumentCommand{\exsub}{s m O{} m}{%
  \IfBooleanT{#1}{\EE_{#2}\nolimits\bra*{#4}}%
  \IfBooleanF{#1}{\EE_{#2}\nolimits\bra[#3]{#4}}%
}

\newcommand{\memo}[1]{{\color{orange}}}

\newcommand{\tbeta}{\bar{\beta}}
\newcommand{\tdelta}{\bar{\delta}}
\newcommand{\sumdelta}[1]{S_{#1}}

\declaretheorem[
  name=Theorem,
  refname={Theorem,Theorems},
  style=thmsty,
]{theorem}

\declaretheorem[
  name=Lemma,
  refname={Lemma,Lemmas},
  style=thmsty,
]{lemma}
\declaretheorem[
  name=Definition,
  refname={Definition,Definitions},
  style=thmsty,
]{definition}

\crefname{algorithm}{Algorithm}{Algorithms}
\crefname{line}{Line}{Lines}
\crefname{section}{Section}{Sections}
\crefname{appendix}{Appendix}{Appendices}
\crefname{table}{Table}{Tables}
\crefname{figure}{Figure}{Figures}
\crefname{equation}{}{}
\Crefname{equation}{Eq.}{Eqs.}

\setlist[itemize]{
  topsep=0.4\baselineskip,
  itemsep=0\baselineskip,
  leftmargin=1.5em,
}

\setlist[enumerate]{
  font=\upshape,
  label=(\alph*),
  ref=(\alph*),
  topsep=0.4\baselineskip,
  itemsep=0\baselineskip,
  leftmargin=2em,
}

\newlist{enuminasm}{enumerate}{1} 
\setlist[enuminasm]{
  font=\upshape,
  label=(\alph*),
  ref=\theassumption(\alph*),
  topsep=0.4\baselineskip,
  itemsep=0\baselineskip,
  leftmargin=2em,
}
\crefalias{enuminasmi}{assumption}

\newlist{enuminthm}{enumerate}{1}
\setlist[enuminthm]{
  font=\upshape,
  label=(\alph*),
  ref=\thetheorem(\alph*),
  topsep=0.4\baselineskip,
  itemsep=0\baselineskip,
  leftmargin=2em,
}
\crefalias{enuminthmi}{theorem}

\newlist{enuminlem}{enumerate}{1}
\setlist[enuminlem]{
  font=\upshape,
  label=(\alph*),
  ref=\thelemma(\alph*),
  topsep=0.4\baselineskip,
  itemsep=0\baselineskip,
  leftmargin=2em,
}
\crefalias{enuminlemi}{lemma}

\newcommand{\email}[1]{\href{mailto:#1}{\nolinkurl{#1}}}
\usepackage{algorithmic}

\usepackage{datetime}
\date{\vspace{-2.5\baselineskip}}

\author[1]{Yasunori Akagi\footnote{Corresponding author. E-mail: \email{yasunori.akagi@ntt.com}}}
\author[1]{Takeshi Kurashima}

\affil[1]{NTT Human Informatics Laboratories, Kanagawa, Japan}

\title{Delta Matters: An Analytically Tractable Model for\\$\beta$\nobreakdash--$\delta$ Discounting Agents}

\begin{document}
\maketitle

\begin{abstract}
Humans exhibit time-inconsistent behavior, in which planned actions diverge from executed actions. Understanding time inconsistency and designing appropriate interventions is a key research challenge in computer science and behavioral economics. Previous work focuses on progress-based tasks and derives a closed-form description of agent behavior, from which they obtain optimal intervention strategies. They model time-inconsistency using the $\beta$--$\delta$ discounting (quasi-hyperbolic discounting), but the analysis is limited to the case $\delta = 1$. In this paper, we relax that constraint and show that a closed-form description of agent behavior remains possible for the general case $0 < \delta \le 1$. Based on this result, we derive the conditions under which agents abandon tasks and develop efficient methods for computing optimal interventions. Our analysis reveals that agent behavior and optimal interventions depend critically on the value of $\delta$, suggesting that fixing $\delta = 1$ in many prior studies may unduly simplify real-world decision-making processes.
\end{abstract}

\section{Introduction}
Human decision making often requires intertemporal choices, and time preference, that is, the extent to which future value is discounted, plays a crucial role \citep{frederick2002time}. In economics, time preference has traditionally been modeled by the exponential discounting framework, which discounts value at a constant rate over time \citep{samuelson1937note}. Under exponential discounting, present and future values are evaluated consistently, yielding \emph{time consistent} behavior: decisions do not change as time elapses.

However, actual human behavior frequently exhibits \emph{time-inconsistency}, in which planned actions diverge from executed actions. 
For example, consider an individual who is planning to diet. At the planning stage, they believe that by foregoing indulgent weekend meals they will succeed in their diet. However, as the weekend approaches, the immediate allure of the indulgent meal intensifies, making it impossible to resist; the individual indulges and consequently fails to adhere to the diet plan.
This shift in valuation between planning and execution is known as time inconsistency, and as the example illustrates, it can impede long-term goal achievement. Consequently, it has become an important research challenge in behavioral economics and computer science to model the mechanisms underlying time-inconsistent behavior mathematically, predict such behavior, and derive optimal interventions that support goal attainment.

One prominent example of this line of research is the Directed Acyclic Graph (DAG) model of time-inconsistent behavior proposed by \citet{kleinberg2014time}. In this model, tasks are represented as a DAG, and an agent repeatedly selects a path that minimizes cost computed under $\beta$--$\delta$ discounting (with $\delta=1$ fixed), thereby reproducing prototypical behaviors such as procrastination and task abandonment while flexibly capturing diverse real-world task structures. Building on this, \citet{akagi2024analytically} introduced an analytically tractable model on progress-based tasks, enabling rigorous analysis of task abandonment conditions and designing interventions that maximize progress.

\begin{figure}[t]
  \centering
  \includegraphics[width=0.5\linewidth]{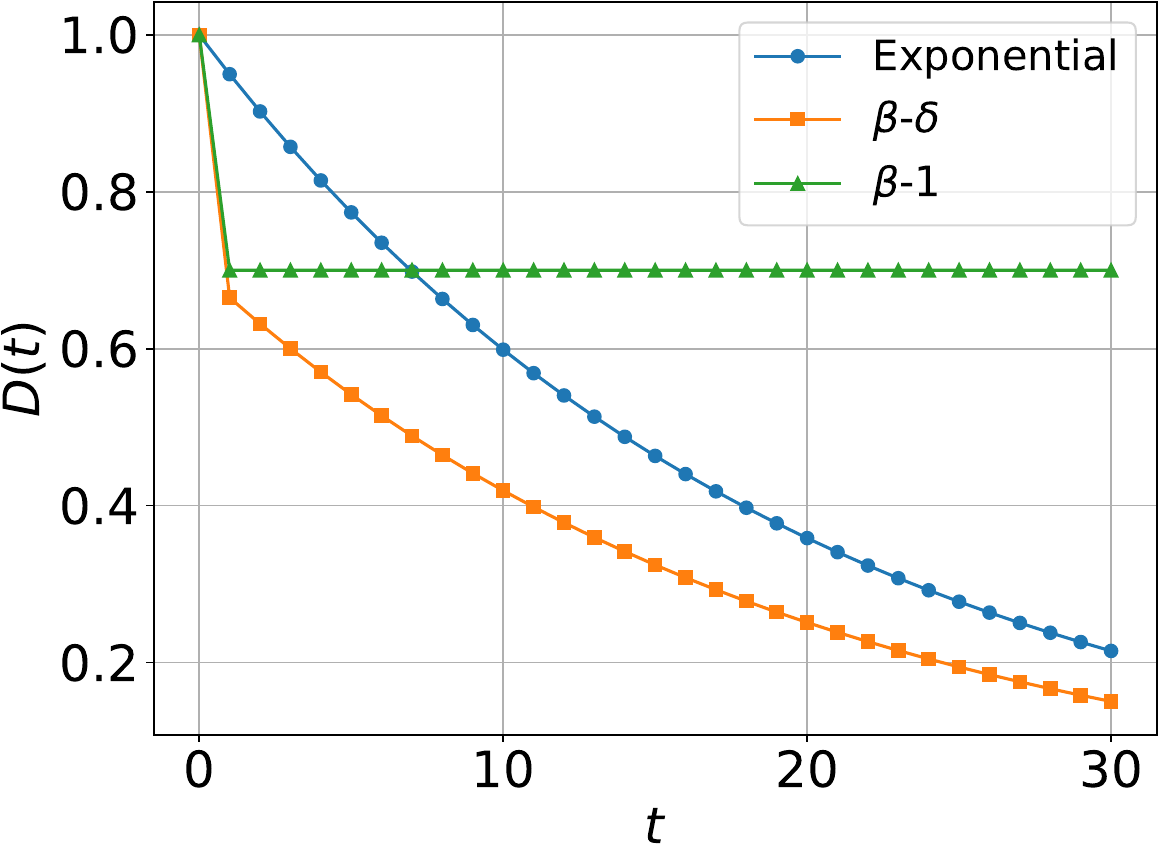}
  \caption{An example of discount functions. We plot the discount rate $D(t)$ at time $t$ for exponential discounting with rate $\delta$, $\beta$-$\delta$ discounting, and $\beta$-1 discounting for $\beta = 0.7$ and $\delta = 0.95$.
}
  \label{fig:discounting example}
\end{figure}

These models represent time preferences using $\beta$–$\delta$ discounting (quasi-hyperbolic discounting) \citep{phelps1968second,laibson1997golden}. Under $\beta$–$\delta$ discounting, a payoff occurring at time $t>0$ is multiplied by the factor $\beta \delta^t$, where $0<\beta,\delta\le1$ are parameters that shape the discount.
The parameter $\beta$ governs the degree of \emph{present bias}, which is the overweighting of immediate value relative to future value, while $\delta$ governs the rate at which future value is further discounted over time, capturing long-term patience.
This discounting scheme contrasts with the classical exponential model, in which the discount factor is given by
$
\delta^t. 
$
\cref{fig:discounting example} illustrates the shapes of the discount functions.

Although $\beta$ and $\delta$ are essential for modeling human time preference, prior work has predominantly focused on the effects of $\beta$, fixing $\delta=1$ to isolate the impact of present bias. For example, \citet{kleinberg2014time} analyze agent behavior under the assumption $\delta=1$, and many subsequent studies have retained this assumption, discussing only the role of $\beta$ \citep{kleinberg2016planning,kleinberg2017planning,akagi2024analytically,tang2017computational,albers2019motivating,gravin2016procrastination}. The assumption $\delta=1$ implies that a reward received after 5 days is valued the same as one received after 365 days, which is clearly unrealistic (see \cref{fig:discounting example}). Nevertheless, the influence of $\delta$ on agent behavior and the optimal design of interventions under $\beta$--$\delta$ discounting has received little attention.

This paper analyzes how the parameter $\delta$ affects agent behavior and optimal interventions. To this end, we build on the analytically tractable model for progress-based tasks introduced by \citet{akagi2024analytically}. First, we extend their model to the general case $\delta \neq 1$ and show that the agent's behavior still admits a closed-form description. Based on this closed-form, we derive theoretically the conditions under which an agent abandons a task and examine which combinations of $(\beta,\delta)$ make abandonment more likely. We then address two optimal intervention problems, goal optimization and reward scheduling, design efficient algorithms to solve them, and use these algorithms to analyze how optimal interventions vary with the parameters $(\beta,\delta)$.

Our analysis yields the following insights:
\begin{itemize}
  \item \emph{Task abandonment.} Whether an agent may abandon a task midway depends not only on $\beta$ but on the combination $(\beta,\delta)$. In particular, a smaller $\delta$ makes abandonment more likely. This contrasts with exponential discounting, where abandonment never occurs for any $\delta$.
  \item \emph{Goal-setting optimization.} In the problem of deciding the optimal goal to maximize final progress, $\delta$ plays a critical role. Notably, exploitative rewards (announced rewards that the agent cannot actually obtain) are most effective when $\beta$ is small and $\delta$ is large. Though powerful, exploitative rewards raise ethical concerns, making our analysis important for guiding their appropriate use.
  \item \emph{Reward-scheduling optimization.} In the problem of optimally timing rewards to maximize overall progress, $\delta$ has a major influence. In realistic parameter ranges, empirically estimated as $\beta\approx$0.5--0.9, $\delta\approx$0.90--0.99 \citep{laibson2024estimating,cheung2021quasi}, the optimal schedule structure is highly sensitive to $\delta$.
\end{itemize}
These results demonstrate that $\delta$ influences agent behavior and optimal interventions in a fundamentally distinct manner from $\beta$, and suggest that fixing $\delta=1$ as in much prior work may excessively simplify real-world decision processes.

All proofs are deferred to the appendix.

\section{Related Work}
Various time discount functions have been proposed to model human time preferences.  Representative examples include exponential discounting\citep{samuelson1937note}, long employed in economics; hyperbolic discounting\citep{ainslie1975specious}, introduced to capture time‐inconsistent preferences; and the $\beta$--$\delta$ discounting (quasi‐hyperbolic discounting), which plays a central role in this study.
Subsequent research has further generalized these discount functions, such as generalized hyperbolic discounting \citep{loewenstein1992anomalies} and generalized Weibull discounting \citep{takeuchi2011non}.

Numerous studies have explored time discounting, time-inconsistency, task abandonment, and optimal intervention. One of the latest pivotal works is DAG model of time-inconsistent behavior proposed by \citet{kleinberg2014time}.
This model has gathered widespread attention for its combination of modeling flexibility (capturing a broad class of real-world tasks) and tractability (leveraging graph-theoretic and algorithmic techniques). 
Subsequent extensions include more complicated biases \citep{kleinberg2016planning,kleinberg2017planning,gravin2016procrastination} and deriving optimal interventions \citep{tang2017computational,albers2019motivating,albers2021value,halpern2023chunking,belova2024guide}. Notably, all of these works fix $\delta=1$ in $\beta$--$\delta$ discounting. 

The work most closely related to ours is the analytically tractable model for present-biased agents in progress-based tasks introduced by \citet{akagi2024analytically}. Their model can be interpreted as a combination of the “cumulative procrastination” framework of \citet{o2006incentives} and the DAG-based agent model of \citet{kleinberg2014time}. It has also been extended to a continuous time model \citep{akagi2025continuous}, providing a versatile application framework. Our principal contribution lies in generalizing this model to the case $\delta \neq 1$ and analytically characterizing how varying $\delta$ influences agent behavior and optimal intervention. For a detailed comparison between the findings of \citet{akagi2024analytically,akagi2025continuous} and ours, see \cref{sec:discussion}.

Although many studies have underscored the importance of the exponential discount rate $\delta$ in the context of exponential discounting, relatively few have focused on the role of $\delta$ within the $\beta$--$\delta$ framework. For example, \citet{meier2008impatience} conducted a field experiment using credit reports and found that $\delta$, more so than $\beta$, strongly correlates with individuals' default behavior and FICO credit scores. Similarly, \citet{burks2012measures} reported that, in a field study of truck driver trainees, including $\delta$ as a predictor improved forecasts of outcomes such as smoking behavior, credit scores, BMI, and job performance, compared to using $\beta$ alone. Although our theoretical results do not directly validate these empirical findings, they share the common direction of highlighting the significance of $\delta$. 
We believe that future work must validate relationships between them through both theoretical analysis and experimental investigation.

\section{Model}
\subsection{Progress-based Task}
This study considers \emph{progress-based tasks} proposed by \citet{akagi2024analytically}. In this type of task, an agent accumulates a quantity called \emph{progress} over a fixed period and receives a reward if a pre-specified progress target is achieved within the time limit. 
Such tasks frequently occur in everyday life. For example, the task ``exercise for 30 hours within one month to improve one's health'' falls into this category; here, the period is one month, and the progress corresponds to the cumulative hours of exercise performed. 
We assume that progress is non-decreasing over time\footnote{In a dieting task where weight loss is progress, that progress can decrease (i.e., weight regain). Tasks in which progress may decline fall outside the scope of this work and are left for future research.}. 

We treat time as a discrete quantity. We denote the total length of the period by $T \in \mathbb{Z}_{>0}$, the goal progress by $\theta \in \mathbb{R}_{>0}$, and the reward by $R \in \mathbb{R}_{>0}$. The agent's state is represented by the tuple $(t, x)$, where $t$ is the time step and $x$ is the progress.

\subsection{Model Definition}
We describe the decision-making model of an agent in a progress-based task.
The proposed model with $\delta = 1$ corresponds to the model proposed by \citet{akagi2024analytically}; thus, the proposed model is a generalization of their model.

An agent in state $(t-1, x_{t-1})$ computes the cost of the states sequence $(t, y_t), \ldots, (T, y_T)$ taken from time $t$ onward as follows:
\begin{align}
  \mathcal{C}_t(y_t, \ldots, y_T)
    &= c(y_t - x_{t-1}) + \sum_{i=t+1}^T \beta \delta^{i-t} c(y_i - y_{i-1}) \\
    &\quad - \beta \delta^{T-t+1} R \cdot \bm{1}[y_T \geq \theta], \label{eq:sequence cost}
\end{align}
where $\bm{1}(\cdot)$ is the indicator function, taking the value 1 if $y_T \ge \theta$ and 0 otherwise, and $c(\Delta)$ denotes the cost required to generate progress $\Delta$. In this study, we assume
\begin{align}
  c(\Delta) =
  \begin{dcases*}
    \Delta^\alpha, & if $\Delta \ge 0$, \\
    +\infty,       & if $\Delta < 0$,
  \end{dcases*}
\end{align}
where $\alpha > 1$ is a parameter determining the shape of the cost function, and $c(\Delta) = +\infty$ for $\Delta < 0$ reflects the model assumption that progress cannot decrease.

\begin{figure}[t]
  \centering
  \includegraphics[width=0.6\linewidth]{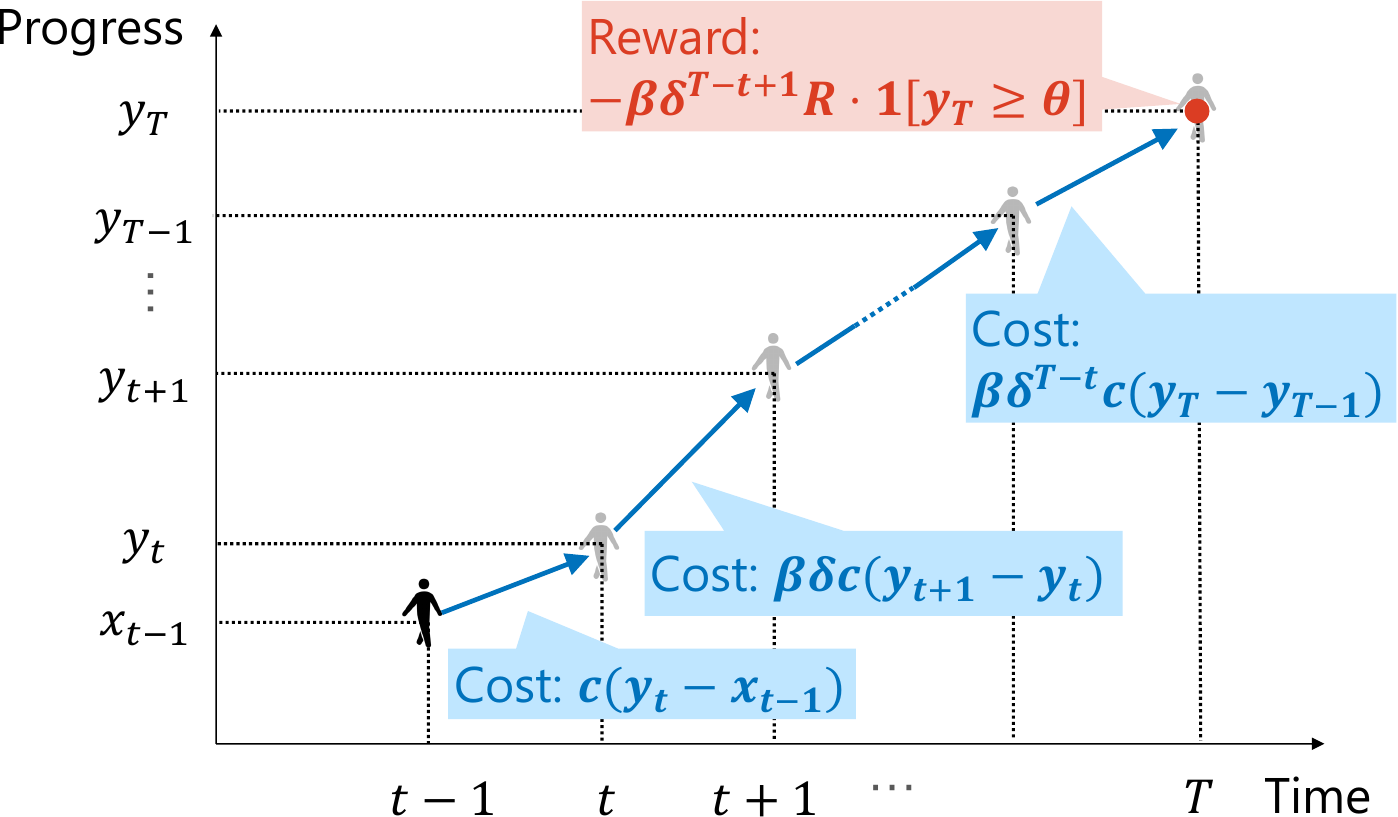}
  \caption{An example of the cost of the states sequence $(t, y_t), \ldots, (T, y_T)$. 
}
  \label{fig:Ct example}
\end{figure}

The cost function $\mathcal{C}_t(y_t, \ldots, y_T)$ is designed according to quasi-hyperbolic discounting~\citep{phelps1968second,laibson1997golden}. In quasi-hyperbolic discounting, the parameters $\beta$ and $\delta$ satisfy $0 < \beta, \delta \le 1$ and determine the form of temporal discounting. When an agent is in state $(t-1, x_{t-1})$, the immediate cost $c(y_t - x_{t-1})$ is not discounted, whereas a cost incurred at time $i\ (> t)$, $c(y_{i} - y_{i-1})$, is discounted by the factor $\beta \delta^{i-t}$. The parameter $\beta$ captures the weight given to all future costs relative to the present, and $\delta$ captures how the discounting of future costs increases over time.
The third term represents the reward: if the agent’s final progress $y_T$ meets or exceeds the goal $\theta$, the agent obtains the discounted reward $\beta \delta^{T-t+1} R$. The preceding minus sign indicates that the reward plays the inverse role of costs.
\cref{fig:Ct example} illustrates an example of computation of the cost of the states sequence $(t, y_t), \ldots, (T, y_T)$. 

At state $(t-1, x_{t-1})$, the agent proceeds as follows:
\begin{itemize}
  \item For each candidate sequence of future states $(t, y_t), \ldots, (T, y_T)$, compute the cost $\mathcal{C}_t(y_t, \ldots, y_T)$ using \cref{eq:sequence cost} and let $(t, y_t^*), \ldots, (T, y_T^*)$ be the sequence that minimizes this cost.
  \item Move from $(t-1, x_{t-1})$ to $(t, y_t^*)$.
\end{itemize}
That is, the sequence of states is determined by $x_0 = 0$ and
\begin{align}
  x_t \coloneqq \argmin_{y_t \in \R} \min_{y_{t+1}, \dots, y_T \in \R} \mathcal{C}_t(y_t, \ldots, y_T) \label{eq:min min}. 
\end{align}
This behavioral model reflects that, at each time step, the agent evaluates the cost of each candidate future sequence under $\beta$--$\delta$ discounting and follows the sequence that yields the minimal cost.

\subsection{Closed Form Progress}

\begin{theorem} \label{thm:state sequence}
  The agent's progress $x_t$ can be written as
  \begin{align}
    x_t = \theta \prn*{1 - \prod_{i=1}^{\min \set*{t, t^*}} p_i},
  \end{align}
  where
  \begin{align}
    p_t &\coloneqq \frac{\sum_{i=1}^{T-t} \tdelta^i}{\sum_{i=1}^{T-t} \tdelta^i + \tbeta\tdelta^{T-t+1}}, \\
    \tbeta &\coloneqq \beta^{\frac{1}{\alpha-1}}, 
    \quad
    \tdelta \coloneqq \delta^{\frac{1}{\alpha-1}}, 
  \end{align}
  and $t^*$ is defined as the smallest $t \in \{0,\dots,T-1\}$ satisfying
  \begin{align}
    \prn*{\sum_{i=1}^{T-t-1} \tdelta^i + \tbeta\tdelta^{T-t}}^{1 - \alpha}
    \prod_{i=1}^t p_i^\alpha
    >
    \frac{R}{\theta^\alpha},
    \label{eq:condition_for_tast}
  \end{align}
  if such a $t$ exists; otherwise, $t^* = T$.
\end{theorem}

\cref{thm:state sequence} indicates that an agent's progress admits a closed-form expression. The quantity $t^*$ corresponds to the time step at which the agent abandons the task, and $t^* = T$ signifies that the agent never abandons the task and completes it. This result extends the $\delta = 1$ case obtained in \cite{akagi2024analytically} to the general case $\delta \leq 1$. Following, we analyze the behavior of $\beta$--$\delta$ agents based on this expression.

\section{Task Abandonment}

This section investigates the relationship between the conditions of task abandonment of agents and the discount parameters. Understanding this relationship enables the prediction of task abandonment and helps design intervention strategies to maximize the agent's final progress. 

First, we extend the concept of the \emph{Task-Abandonment Inducing (TAI)} parameter proposed in \citep{akagi2024analytically} to the case of $\beta$--$\delta$ discounting.

\begin{definition}
For a $\beta$--$\delta$ agent, if there exist $\theta,R \in \mathbb{R}_{\ge 0}$ such that $t^* \neq 0$ and $t^* \neq T$, then the discount parameters $(\beta,\delta)$ are said to be \emph{Task-Abandonment Inducing} (TAI).
\end{definition}

Here, $t^*$ is the time step at which the agent abandons the task, as defined in \cref{thm:state sequence}. If $t^* = 0$, the agent abandons the task from the outset, i.e., never begins. If $t^* = T$, the agent never abandons and completes the task. Hence, an agent with TAI parameters may abandon the task midway for some $\theta$ and $R$. Conversely, an agent without TAI parameters will never abandon midway for any $\theta$ or $R$, and will either never start or see the task through to completion. The TAI property thus serves not only as an indicator of propensity for task abandonment, but also has significant implications for optimal intervention strategies (see \cref{sec:optimal goal setting,sec:optimal scheduling}). 

Define
\begin{align}
  q_t \coloneqq \prn*{\sum_{i=1}^{T-t-1} \tdelta^i + \tbeta\tdelta^{T-t}}^{1 - \alpha}
  \prod_{i=1}^t p_i^\alpha,
  \label{eq:def qt}
\end{align}
which is the left-hand side of \cref{eq:condition_for_tast}. By \cref{thm:state sequence}, $t^*$ is the smallest $t$ satisfying $q_t > \frac{R}{\theta^\alpha}$. Therefore, the necessary and sufficient condition for $(\beta,\delta)$ to be TAI is
\begin{align}
  \max_{t \in \{0,1,\ldots,T-1\}} q_t \neq q_0.
  \label{eq:TAI q_t condition}
\end{align}

We derive the necessary and sufficient condition for $(\beta, \delta)$ to be TAI. 
To present this result, we prepare some preliminaries. First, define
\begin{align}
  h_1(x) &\coloneqq \prn*{\frac{(\alpha - 1)\prn*{1 - x^{\frac{1}{\alpha - 1}}}}{x^{\frac{1}{\alpha - 1}}}}^{\alpha - 1}, \\
  h_2(x) &\coloneqq \prn*{\frac{(\alpha - 1)\prn*{1 - x^{\frac{1}{\alpha - 1}}}}{x^{\frac{1}{\alpha - 1}}\prn*{1 - \alpha(1 - x^{\frac{1}{\alpha - 1}})}}}^{\alpha - 1}, 
  \\ \gamma_1 &\coloneqq \prn*{1 - \tfrac{1}{\alpha}}^{\alpha - 1}, 
  \quad
  \gamma_2 \coloneqq \prn*{1 - \tfrac{1}{\alpha}}^{\frac{\alpha - 1}{2}}.
\end{align}
Then we have:
\begin{enumerate}
  \item $h_1(\gamma_1) = 1$, $h_2(\gamma_2) = 1$, and $h_1(1) = h_2(1) = 0$,
  \item $h_1$ is strictly decreasing on $[\gamma_1,1]$, and $h_2$ is strictly decreasing on $[\gamma_2,1]$.
\end{enumerate}
Hence $h_1$ and $h_2$ admit inverse functions on $[\gamma_1,1]$ and $[\gamma_2,1]$, respectively; denote these inverses by $h_1^{-1}$ and $h_2^{-1}$. With these preparations, we state the following theorem.

\begin{theorem}\label{thm:TAI condition}
  Fix $\delta\in(0,1]$. Then there exists a threshold $\beta_0(\delta)$ satisfying
  $
    h_1^{-1}(\delta) < \beta_0(\delta) < h_2^{-1}(\delta),
  $
  and $(\beta,\delta)$ is TAI if and only if $\beta < \beta_0(\delta)$.
\end{theorem}

\cref{thm:TAI condition} implies that for a fixed $\delta$, the TAI property is determined by whether $\beta$ is smaller than the threshold $\beta_0(\delta)$. Although deriving an analytic form for $\beta_0(\delta)$ is difficult, it is guaranteed to lie between $h_1^{-1}(\delta)$ and $h_2^{-1}(\delta)$. Note that these inverse functions can be expressed explicitly as functions of $\delta$ (see \cref{sec:properties of h_inv}). 

\begin{figure}[t]
  \centering
  \includegraphics[width=0.5\linewidth]{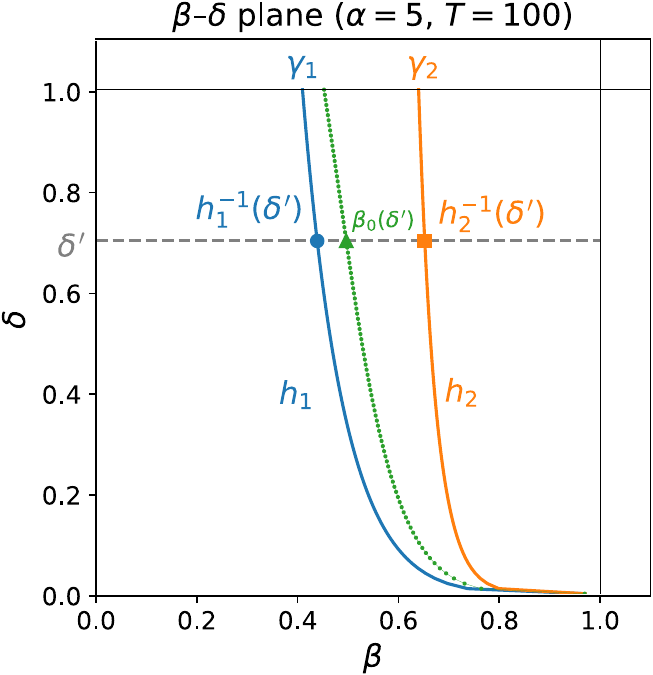}
  \caption{The relationship among $h_1$, $h_2$, and $\beta_0$ in the $\beta$--$\delta$ plane for $\alpha = 5$ and $T = 100$. The green dashed line shows the numerically computed values of $\beta_0(\delta)$.}
  \label{fig:beta delta plane}
\end{figure}

\begin{figure}[t]
  \begin{minipage}[b]{0.45\linewidth}
    \centering
    \includegraphics[width=\linewidth]{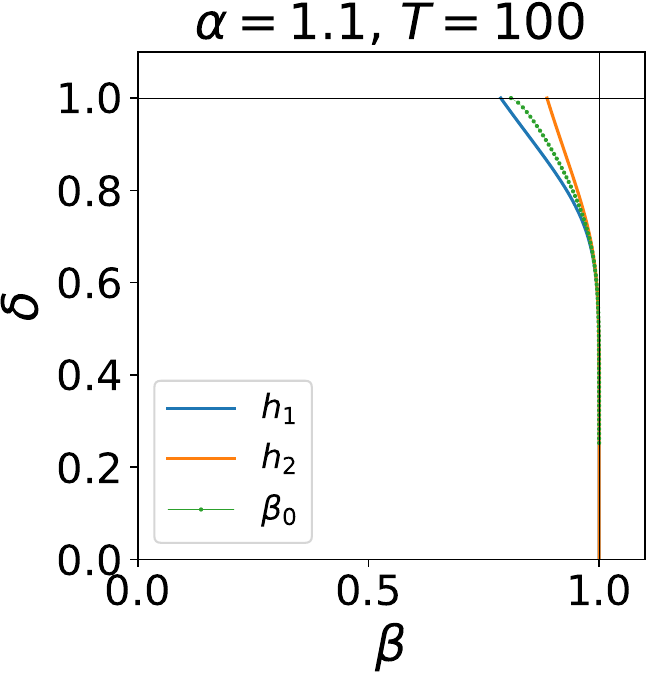}
  \end{minipage}
  \hfill
  \begin{minipage}[b]{0.45\linewidth}
    \centering
    \includegraphics[width=\linewidth]{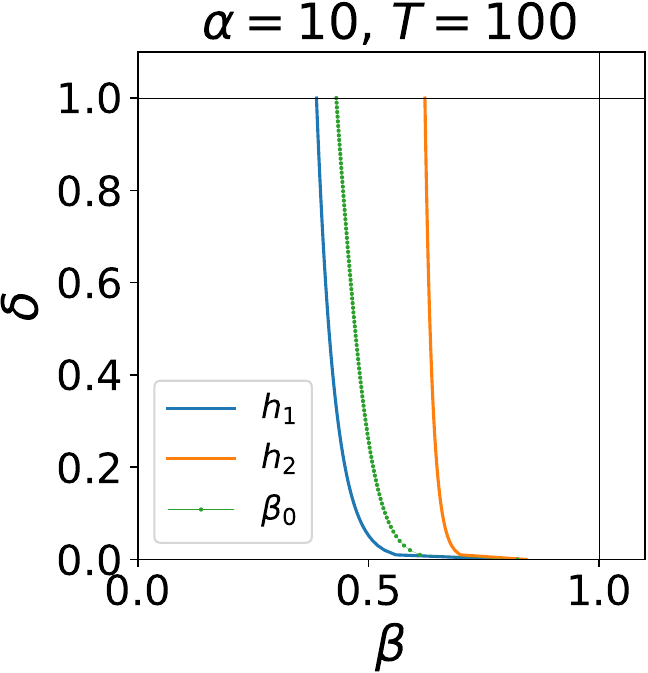}
  \end{minipage}
  \caption{Left: the relationship among $h_1$, $h_2$, and $\beta_0$ for $\alpha = 1.1$, $T = 100$; Right: the same for $\alpha = 10$, $T = 100$.}
  \label{fig:beta delta plane simple}
\end{figure}

\cref{fig:beta delta plane} illustrates the relationship among $h_1$, $h_2$, and $\beta_0$ in the $\beta$--$\delta$ plane for the case $\alpha = 5$ and $T = 100$. The green dashed line plots the numerically computed $\beta_0(\delta)$.\interfootnotelinepenalty=10000
\footnote{By \cref{lem:q_beta_closer} in \cref{sec:proof of thm:TAI condition}, $\beta_0(\delta)$ is the value of $\beta$ satisfying $q_0 = q_{T-1}$. Because \cref{lem:q_beta_closer} guarantees that $q_0<q_{T-1}$ for $h_1^{-1}(\delta)\le\beta<\beta_0(\delta)$ and $q_0>q_{T-1}$ for $\beta_0(\delta)<\beta\le h_2^{-1}(\delta)$, for fixed $\delta$, we can find $\beta_0(\delta)$ numerically via binary search. }
Moreover, \cref{fig:beta delta plane simple} plots the same relationship for $\alpha = 1.1, T=100$ and $\alpha = 10, T=100$. 

From \cref{fig:beta delta plane}, we observe:
\begin{itemize}
  \item The $\beta$--$\delta$ space is divided into two connected regions: one where the parameters are TAI and one where they are not.
  \item Smaller values of $\beta$ tend to induce the TAI property.
  \item Smaller values of $\delta$ also tend to induce the TAI property.
\end{itemize}

Furthermore, \cref{fig:beta delta plane simple} indicates that the TAI region expands as $\alpha$ decreases. This observation is supported by
$
  \lim_{\alpha \to 1^+} h_1^{-1}(\delta) = \lim_{\alpha \to 1^+} h_2^{-1}(\delta) = 1, 
  \lim_{\alpha \to \infty} h_1^{-1}(\delta) = \frac{1}{e}, 
  \lim_{\alpha \to \infty} h_2^{-1}(\delta) = \frac{1}{\sqrt{e}}
$
for $0 < \delta \le 1$,
where $e$ is Euler's number (see \cref{sec:properties of h_inv} for proofs). Intuitively, when $\alpha$ is small, the cost of generating a large amount of progress in a short time is low, making the strategy of “catching up at the end” more attractive and thus increasing the tendency to procrastinate, resulting in task abandonment. 

These results demonstrate that whether the parameter pair is TAI depends critically not only on $\beta$ but also on $\delta$. Therefore, analyses of agent behavior should pay careful attention to both discount parameters.

\section{Goal Optimization}
\label{sec:optimal goal setting}
This section examines how to determine the goal $\theta$ to maximize the agent's final progress $x_T$, given the period $T$ and reward $R$. Concretely, we consider the optimization problem
\begin{align}
  \max_{\theta \geq 0}\ x_T.
\end{align}

This problem differs fundamentally depending on whether \emph{exploitative} rewards are permitted. 
An exploitative reward refers to a decoy incentive the agent can never obtain, since the goal is inherently unreachable.
A time-inconsistent agent may initially believe the goal is attainable and accumulate progress, only to abandon the task later. In such cases, although no reward is paid, the agent accumulates partial progress toward the goal. One can induce the agent to work without ever paying the reward, and in some settings, exploitative rewards can yield greater final progress than non-exploitative schemes. Since exploitative rewards raise ethical concerns, caution is required in practice. This study analyzes the optimal goal-setting problem when exploitative rewards are allowed and when not. Mathematically, disallowing exploitative rewards corresponds to adding the constraint $x_T \ge \theta$ to the original optimization problem.

\begin{theorem}[Non-Exploitative Case]\label{thm:optimal_goal_non_exploitative}
  Suppose exploitative rewards are \emph{not} permitted.
  \begin{enuminthm}
    \item If $(\beta,\delta)$ is \emph{not} TAI, then the optimal goal is
    \begin{align}
      \theta
      = \prn*{\frac{R}{q_{0}}}^{\frac{1}{\alpha}}
      = R^{\frac{1}{\alpha}}\,\prn*{\sum_{i=1}^{T-1}\tdelta^i + \tbeta\,\tdelta^{T}}^{\frac{\alpha-1}{\alpha}},
      \label{eq:optimal_goal_non_TAI}
    \end{align}
    and the final progress achieved $x_T = \theta$.
    \item If $(\beta,\delta)$ \emph{is} TAI, then the optimal goal is
    \begin{align}
      \theta
      = \prn*{\frac{R}{q_{T-1}}}^{\frac{1}{\alpha}}
      = R^{\frac{1}{\alpha}}\,\beta^{\frac{1}{\alpha}}\delta^{\frac{1}{\alpha}}
        \prod_{t=1}^{T-1} \prn*{1 + \frac{\tbeta\,\tdelta^{T-t+1}}{\sum_{i=1}^{T-t}\tdelta^i}},
      \label{eq:optimal_goal_TAI}
    \end{align}
    and again $x_T = \theta$.
  \end{enuminthm}
\end{theorem}

\begin{theorem}[Exploitative Case]\label{thm:optimal_goal_exploitative}
  Suppose exploitative rewards \emph{are} permitted.
  \begin{enuminthm}
    \item If $(\beta,\delta)$ is \emph{not} TAI, then the optimal goal remains
    \begin{align}
      \theta
      = \prn*{\frac{R}{q_{0}}}^{\frac{1}{\alpha}}
      = R^{\frac{1}{\alpha}}\,\prn*{\sum_{i=1}^{T-1}\tdelta^i + \tbeta\,\tdelta^{T}}^{\frac{\alpha-1}{\alpha}},
      \label{eq:optimal_goal_non_TAI_exploitative}
    \end{align}
    as in the non-exploitative case, and $x_T = \theta$.
    \item\label{thm:optimal_goal_TAI_exploitative}
    If $(\beta,\delta)$ \emph{is} TAI, then the optimal goal is
    \begin{align}
      \theta
      = \prn*{\frac{R}{\max\{q_0,\,q_{t^*-1}\}}}^{\frac{1}{\alpha}},
      \label{eq:optimal_goal_TAI_exploitative}
    \end{align}
    where
    \begin{align}
      t^* &\coloneqq \argmax_{t\in\{\tilde t,\dots,T\}} u_t,
      \label{eq:t^*}\\
      u_t &\coloneqq \prn*{\frac{R}{\max\{q_0,\,q_{t-1}\}}}^{\frac{1}{\alpha}}
              \prn*{1 - \prod_{i=1}^t p_i},
    \end{align}
    and $\tilde t$ is the smallest $t$ such that $q_t > q_0$. In this case, $x_T = u_{t^*}$.
  \end{enuminthm}
\end{theorem}

The right-hand sides of 
\eqref{eq:optimal_goal_non_TAI}, \eqref{eq:optimal_goal_TAI}, and \eqref{eq:optimal_goal_non_TAI_exploitative}
are strictly increasing in $\delta$. Hence, in these cases, a smaller $\delta$ calls for a smaller optimal goal $\theta$, resulting in a smaller final progress.

\begin{figure}[t]
  \centering
  \includegraphics[width=0.5\linewidth]{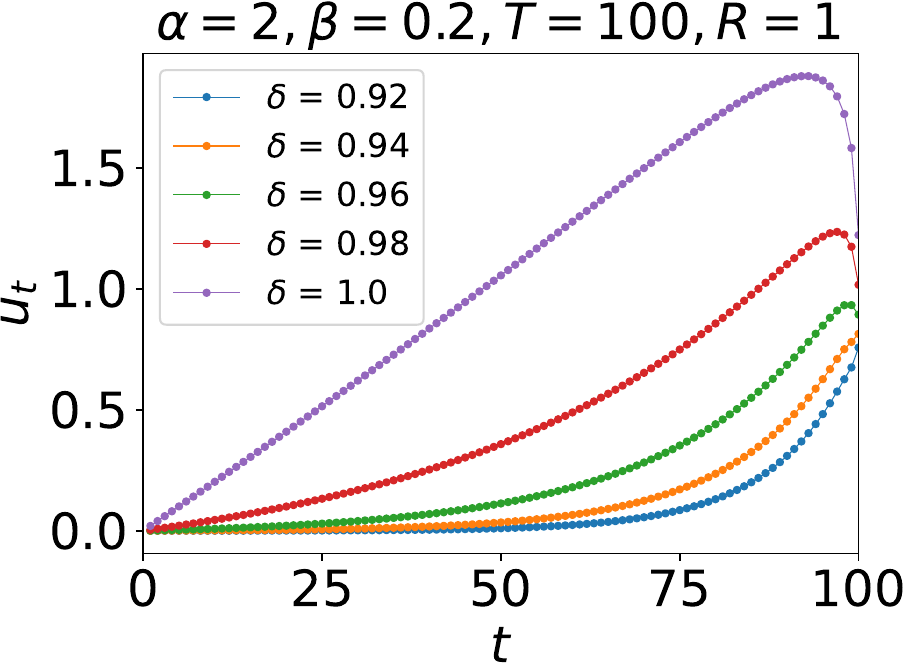}
  \includegraphics[width=0.5\linewidth]{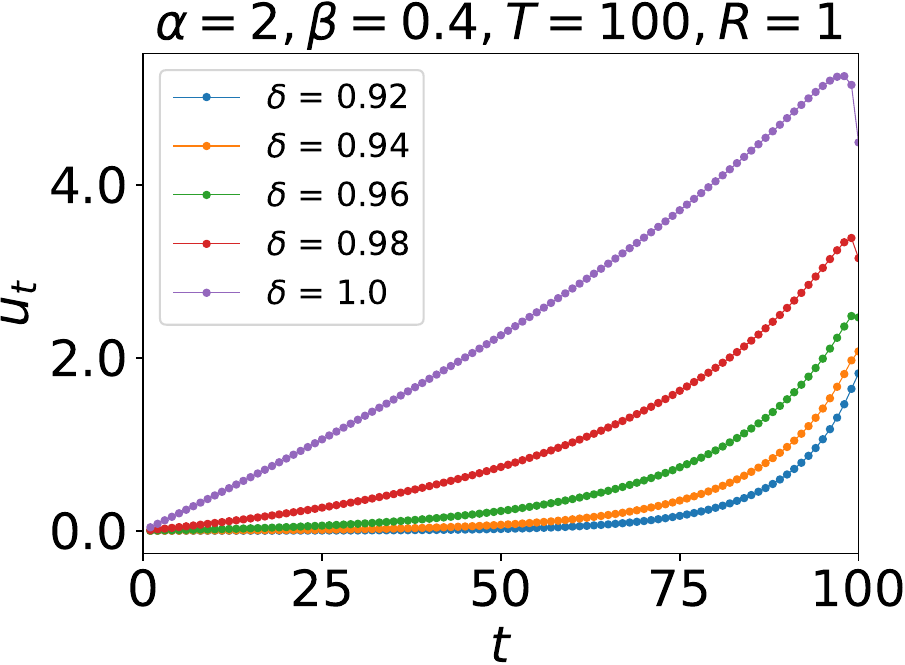}
  \caption{Relationship between $t$ and $u_t$ when $\alpha=2$, $T=100$, and $R=1$ are fixed, and $\beta,\delta$ vary. Top: $\beta=0.2$; bottom: $\beta=0.4$.}
  \label{fig:optimal_goal_TAI_exploitative}
\end{figure}

In the case where the parameters $(\beta,\delta)$ are TAI and exploitative rewards are permitted (\cref{thm:optimal_goal_TAI_exploitative}), the optimal goal is determined by the optimum solution of \cref{eq:t^*}, making it challenging to derive qualitative properties directly from the formula. To explore these properties empirically, we plot the values of $u_t$ for several representative cases in \cref{fig:optimal_goal_TAI_exploitative}.

From \cref{fig:optimal_goal_TAI_exploitative}, we observe the following:
\begin{enumerate}
  \item As $\delta$ decreases, $u_{t^*}$ (where $t^*$ is defined in \eqref{eq:t^*}) decreases; that is, the maximum final progress achievable becomes smaller.
  \item As $\delta$ increases, the gap between the maximum $u_{t^*}$ and $u_T$ widens. Since $u_{t^*}$ is the maximum progress under exploitative rewards and $u_T$ is the maximum progress when exploitative rewards are disallowed, a larger $\delta$ amplifies the exploitative-reward effect.
\end{enumerate}

\begin{figure}[t]
  \centering
  \includegraphics[height=0.435\linewidth]{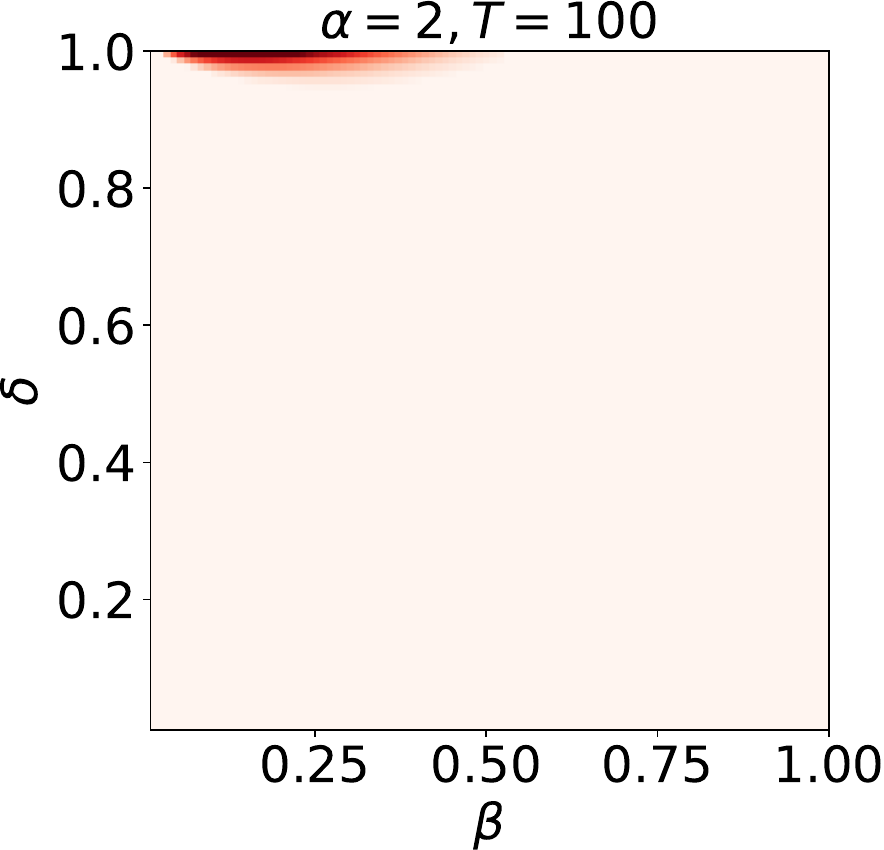}
  \includegraphics[height=0.435\linewidth]{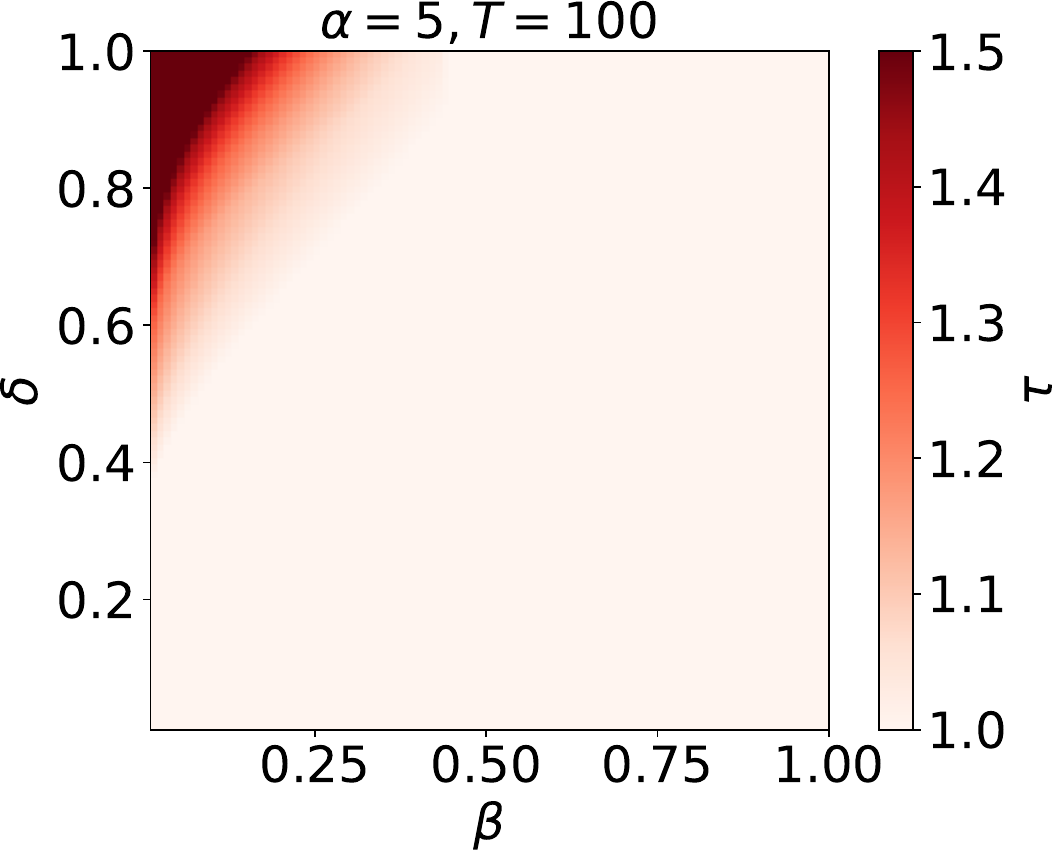}
  \caption{Heatmap of $\tau \coloneqq \frac{u_{t^*}}{u_T}$ on the $(\beta,\delta)$ plane. Darker colors indicate a larger effect of exploitative reward.}
  \label{fig:u_ratio}
\end{figure}

To examine point~(b) in more detail, we define
$
  \tau \coloneqq \frac{u_{t^*}}{u_T},
$
and for fixed $\alpha = 2, 5$ and $T = 100$, we display $\tau$ over the $(\beta,\delta)$ plane as a heat map in \cref{fig:u_ratio}. Since $\tau$ measures the ratio of maximum progress with exploitative rewards to that without, it indicates the effectiveness of exploitative rewards. Larger $\tau$ implies a stronger exploitative effect. \cref{fig:u_ratio} suggests that the exploitative effect is most pronounced when $\beta$ is small and $\delta$ is large.

These observations indicate that $\beta$ and $\delta$ exert opposing influences on the effectiveness of exploitative rewards. We can interpret this phenomenon as follows. For exploitative rewards to be effective, the agent must be motivated by anticipating a reward that is both difficult to obtain and temporally distant. A small $\beta$ (strong present bias) makes the agent prone to procrastination, leading to sustained partial progress toward an unattainable goal and thus accumulating considerable progress for exploitative reward. In contrast, a small $\delta$ (steep long-term discount) diminishes the agent's valuation of distant future rewards, reducing its incentive to accumulate progress and weakening the exploitative effect. 

These insights are crucial to design interventions: we must distinguish between $\beta$ and $\delta$, measure them accurately, and tailor interventions accordingly; otherwise, interventions may be ineffective or unintentionally exploitative.

\section{Reward Scheduling Optimization}
\label{sec:optimal scheduling}
Next, we consider an optimization problem that aims to maximize the agent's final progress by splitting the given total reward and appropriately presenting it to the agent. 
This problem is an extension of the goal-optimization problem in \cref{sec:optimal goal setting}, allowing for more flexible interventions. We call this the \emph{optimal reward scheduling problem}.

\begin{figure}[t]
  \centering
  \includegraphics[width=0.6\linewidth]{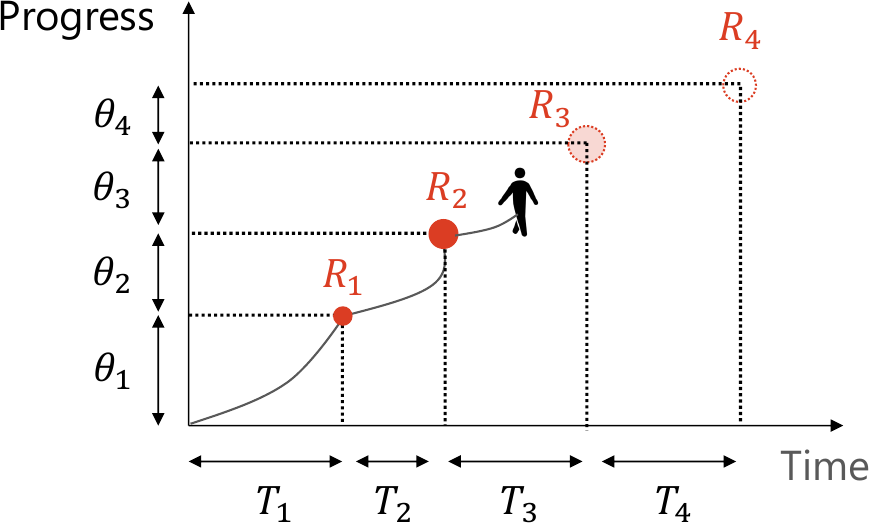}
  \caption{An example of reward scheduling with $N=4$. 
}
  \label{fig:scheduling_N4}
\end{figure}

We mathematically formulate this problem. We assume that exploitative rewards are disallowed. First, the intervener is given a total period $T \in \Z_{>0}$ and a total reward $R \in \R_{\ge0}$. The intervener splits these into $N \in \Z_{>0}$ parts, with subperiods $T_1,\dots,T_N \in \Z_{>0}$ and subrewards $R_1,\dots,R_N \in \R_{\ge0}$. In the $i$-th period, the agent accumulates progress under the condition ``increasing progress by $\theta_i \in \R_{\ge0}$ within period $T_i$ yields reward $R_i$''. The interventor's objective is to determine $N$, $(T_i)_{i=1}^N$, $(R_i)_{i=1}^N$, and $(\theta_i)_{i=1}^N$ so as to maximize the sum of progress across all periods. Note that rewards are presented sequentially: while working in period $i$, the agent does not observe any future rewards or goals for periods $i+1,\dots,N$. 
\Cref{fig:scheduling_N4} illustrates an example of reward scheduling with $N=4$. 

\begin{theorem}\label{thm:optimal_reward_scheduling_non_TAI}
  For given $(\beta, \delta)$, define 
  \begin{align}
    F(x) \coloneqq
    \begin{cases}
    \sum_{t=1}^{x-1} \tdelta^t + \tbeta\,\tdelta^x & \text{$(\beta, \delta)$ is not TAI}, \\
    \tbeta \tdelta \prod_{t=1}^{x-1} \prn*{1 + \frac{\tbeta\,\tdelta^{x-t+1}}{\sum_{i=1}^{x-t}\tdelta^i}}^{\frac{\alpha}{\alpha-1}} & \text{$(\beta, \delta)$ is TAI}. 
    \end{cases}
  \end{align}
  Let $N,(T_i)_{i=1}^N$ be the optimal solution to
  \begin{align}
    \max_{N,(T_i)_{i=1}^N} \ 
    \sum_{i=1}^N F(T_i), \ 
    \st\ 
    \sum_{i=1}^N T_i = T, \ T_i \in \Z_{> 0}, 
    \label{eq:problem_reward_scheduling_reduced}
  \end{align}
  and denote its optimum by $N^*,(T_i^*)_{i=1}^{N^*}$. Then the optimal reward schedule is
  \begin{align}
    N = N^*, \ 
    T_i = T_i^*, \ 
    R_i \propto F(T_i), \ 
    \theta_i = R_i^{\frac{1}{\alpha}} F(T_i)^{\frac{\alpha-1}{\alpha}}.
  \end{align}
\end{theorem}

\begin{figure}[t]
    \includegraphics[width=0.6\linewidth]{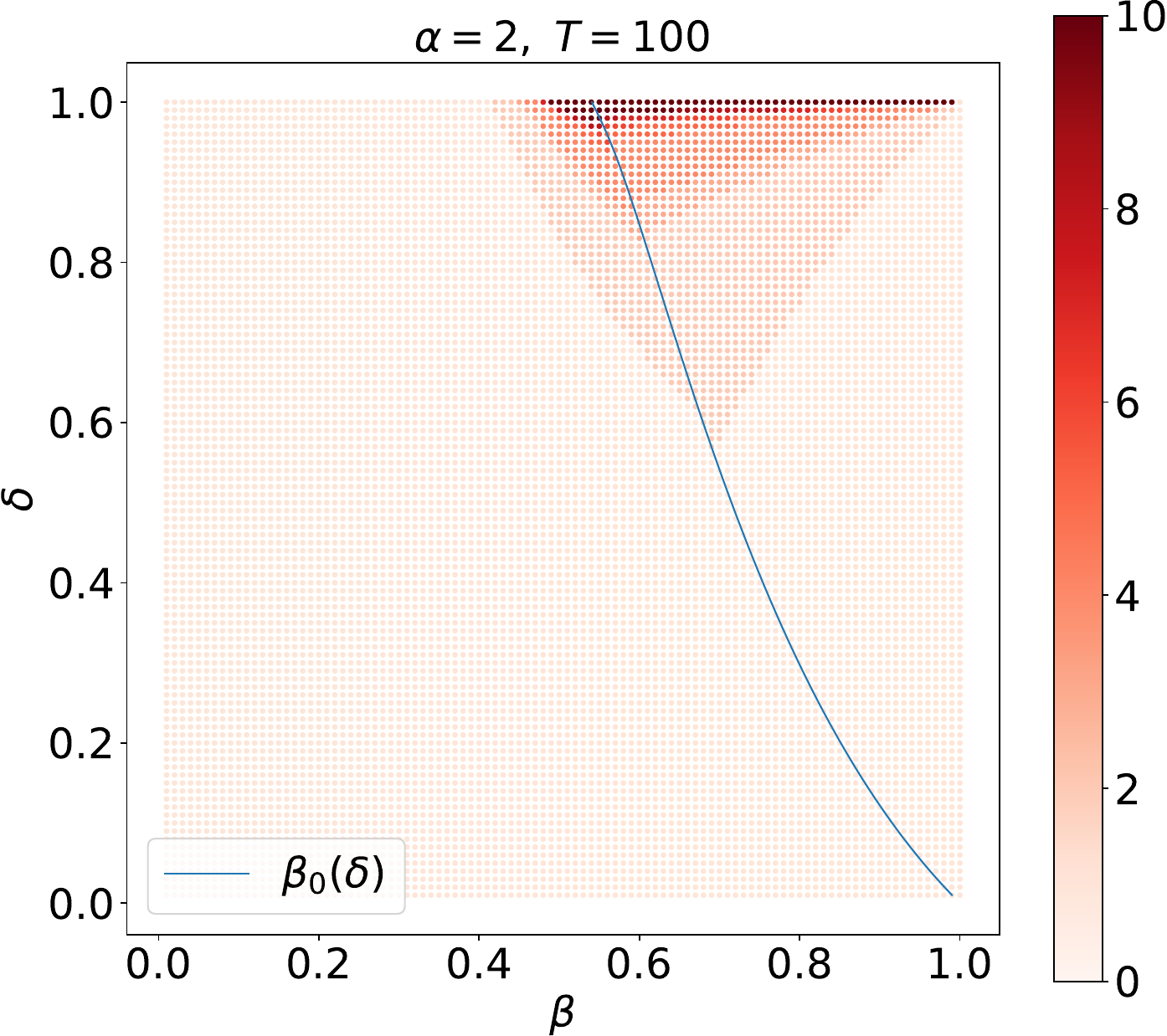}
    \centering
    \includegraphics[width=0.6\linewidth]{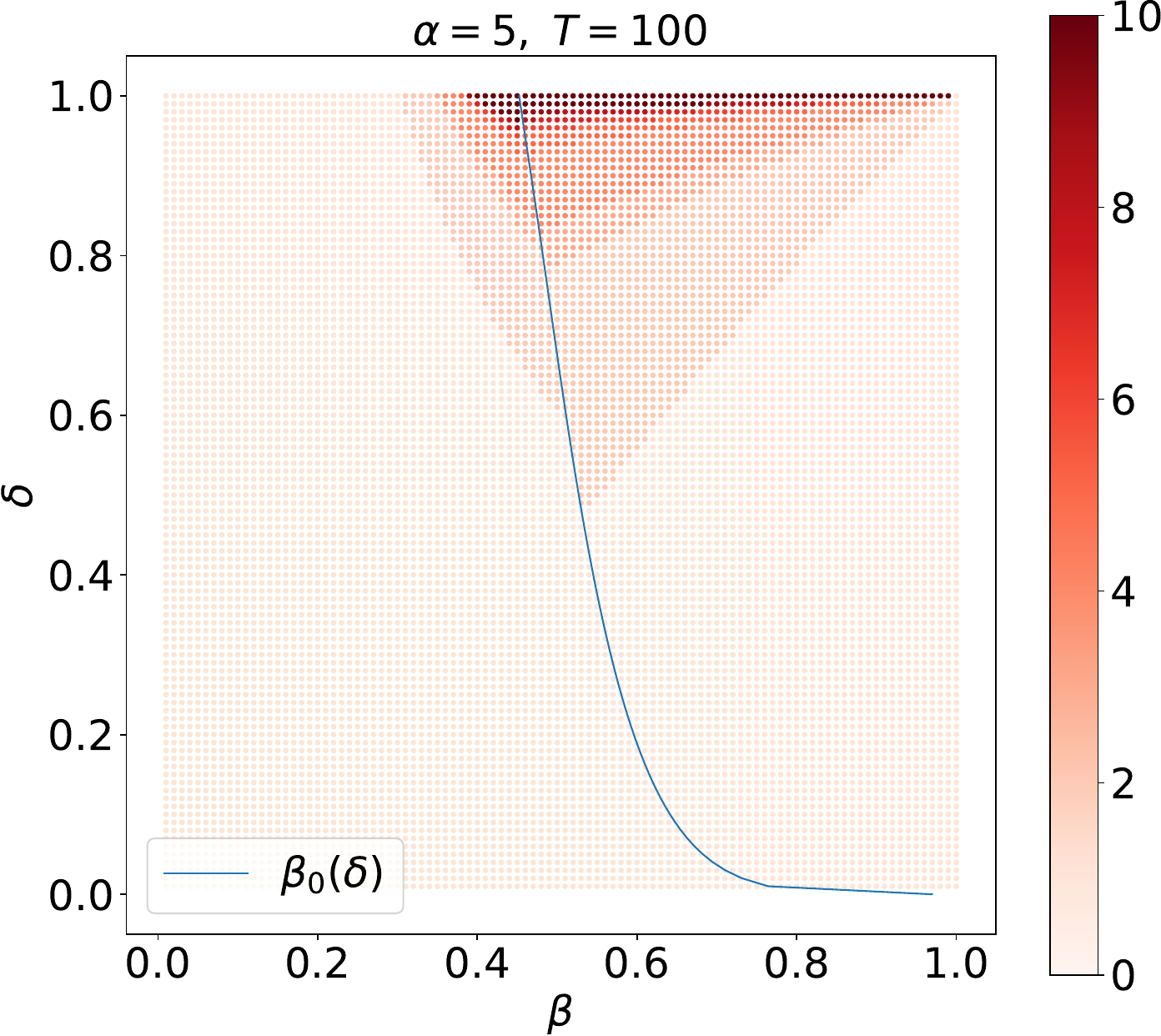}
  \caption{Top: heatmap of $\max_{i=1}^N T_i$ for $\alpha=2, T=100$. Bottom: for $\alpha=5, T=100$. The blue curve denotes the TAI boundary $\beta_0(\delta)$.}
  \label{fig:optimal reward scheduling beta delta plane}
\end{figure}

The optimization problems \eqref{eq:problem_reward_scheduling_reduced} can be solved exactly in $O(T^2)$ time via dynamic programming (see \cref{sec:reward-scheduling-dp} for detail), so we can solve the original reward scheduling optimization problem in $O(T^2)$ time. For $T=100$, we computed the optimal schedule over the grid $\{(\beta,\delta) = (0.01i,0.01j)\mid i,j=1,\dots,100\}$ by dynamic programming and plotted $\max_{i=1}^N T_i$ in Figure~\ref{fig:optimal reward scheduling beta delta plane}. In our computations, we observed $\max_{i=1}^N T_i - \min_{i=1}^N T_i \le 1$ for all $(\beta,\delta)$, so we can regard $\max_{i=1}^N T_i$ as the optimal reward-interval length. From Figure~\ref{fig:optimal reward scheduling beta delta plane}, we see:
\begin{itemize}
  \item For fixed $\beta$, smaller $\delta$ leads to shorter optimal intervals.
  \item Fixing $\delta$ and varying $\beta$: for $\alpha=2$, when $\delta \leq 0.6$, the optimal interval is 1 (i.e.\ every time step), regardless of $\beta$. Otherwise, as $\beta$ increases from 0, the optimal interval first lengthens, then shortens; the maximal interval occurs near the TAI boundary $\beta_0(\delta)$. A similar pattern holds for $\alpha=5$.
\end{itemize}

These results further indicate the importance of $\delta$ in intervention design. Even for the same $\beta$, variations in $\delta$ change the optimal intervention. In particular, in the human's parameter range $\beta\approx$0.5--0.9, $\delta\approx$0.90--0.99 estimated in field studies~\citep{laibson2024estimating,cheung2021quasi}, optimal solution is very sensitive to the value of $\delta$. Therefore, accurate measurement of $\delta$ is crucial for effective intervention design.

\section{Comparison to Previous Models} \label{sec:discussion}
This section discusses the relationships and differences among the insights gained from \citep{akagi2024analytically}, \citep{akagi2025continuous}, and this work. 
The model of \citet{akagi2024analytically} corresponds to fixing $\delta = 1$ in our framework, and thus can be viewed as a special case of our model. The model of \citet{akagi2025continuous} is a continuous-time model and employs hyperbolic discounting rather than $\beta$--$\delta$ discounting. We refer to our model as the \emph{$\beta$--$\delta$ model}, to \citet{akagi2024analytically}'s as the \emph{$\beta$--1 model}, and to \citet{akagi2025continuous}'s as the \emph{continuous-time hyperbolic model (CTHM)}.

\paragraph{Task Abandonment}
The $\beta$--1 model has a threshold $\beta_0$ such that the discount parameters are TAI if and only if $\beta < \beta_0$ \citep[Theorem 2]{akagi2024analytically}. In the $\beta$--$\delta$ model, the analogous generalization (Theorem~\ref{thm:TAI condition} in this paper) holds: $\beta_0$ depends on $\delta$ and tends to increase as $\delta$ decreases. In the 
CTHM
, steep discounting induces task abandonment \citep[Proposition 1]{akagi2025continuous}, displaying the same trend as other models.

\paragraph{Goal-Setting Optimization}
Under the $\beta$--1 model, exploitative rewards become more effective as $\beta$ decreases. The $\beta$--$\delta$ model exhibits a similar pattern, but exploitative rewards are effective only when $\delta$ is sufficiently large. In contrast, the 
CTHM
demonstrates that exploitative rewards are never beneficial: non-exploitative goals always produce higher final progress. This highlights a fundamental divergence between discrete-time models and the
CTHM
.

\paragraph{Reward-Scheduling Optimization}
In the $\beta$--1 model, large $\beta$ favors lump-sum rewards, whereas small $\beta$ favors splitting rewards. The $\beta$--$\delta$ model also requires parameter-dependent reward splitting, but the mapping from $(\beta,\delta)$ to the optimal splitting is more intricate, as in Figure~\ref{fig:optimal reward scheduling beta delta plane}. By contrast, in the 
CTHM
, fine-grained reward subdivision is optimal regardless of parameter values. Here, the discrete-time and continuous-time frameworks also yield markedly different optimal structures.

In summary, while discrete-time models (both $\beta$--1 and $\beta$--$\delta$) and the 
CTHM
 share similar behavior regarding task abandonment, they diverge completely in their optimal solutions for goal-setting and reward scheduling. 
The precise reason for these differences remains unclear. One possible explanation lies in the differing shapes of the discount functions. In hyperbolic discounting used in the 
CTHM
, the discount function is given by
$
D(t) = \frac{1}{1 + k t},
$
where the single positive parameter $k$ determines both the short-term and long-term discount rates, by contrast, in the $\beta$--$\delta$ model, $\beta$ governs short-term discounting and $\delta$ governs long-term discounting independently, thereby allowing a wider variety of discounting behaviors than hyperbolic discounting. This greater flexibility may introduce more complex effects on the optimal intervention strategy under $\beta$--$\delta$ preferences. 
More detailed theoretical analyses may clarify this issue in future work. Additionally, because \citet{akagi2025continuous} has not undertaken empirical validation, assessing experimentally which model more accurately captures real-world human behavior and yields superior intervention strategies is important.
Various techniques to estimate parameters of the $\beta$–$\delta$ model have been established \cite{laibson2024estimating,cheung2021quasi}, and by comparing the estimated parameters with observed human behavior and responses to interventions, such validation can be carried out.

\section{Conclusion} \label{sec:conclusion}
This study extends the time-inconsistent agent behavior model proposed by \citet{akagi2024analytically} to the case $\delta \neq 1$, successfully deriving a closed-form mathematical description of the agent's behavior. 
Based on this expression, we characterize the conditions of task abandonment, derive optimal intervention algorithms, and analyze the relationship between $\delta$ and the optimal interventions.
Our results demonstrate that $\delta$ plays a critical role in agent behaviors and the structure of optimal interventions, suggesting that, unlike prior work which fixes $\delta=1$, this assumption may constitute an undue simplification of real-world decision processes.

\bibliographystyle{abbrvnat}
\bibliography{main}

\clearpage
\appendix
\section{Proof of \cref{thm:state sequence}} \label{sec:proof of thm:state sequence}
\begin{proof}
For simplicity, let
\begin{align}
    \sumdelta{t} &\coloneqq \sum_{i=1}^t \delta^{\frac{i}{\alpha-1}} =  \sum_{i=1}^t \tdelta^i,
\end{align}
and, as defined in \cref{thm:state sequence}, let $\tbeta \coloneqq \beta^{\frac{1}{\alpha-1}}$ and $\tdelta \coloneqq \delta^{\frac{1}{\alpha-1}}$.

Consider the inner minimization in \cref{eq:min min}. If $y_T < \theta$, the minimum is clearly attained at $y_t = y_{t+1} = \cdots = y_T$, yielding an objective value of $0$. Otherwise, the minimum is attained at $y_T = \theta$. Hence,
\begin{align}
\min_{y_{t+1}, \ldots, y_T \in \R} \mathcal{C}_t(y_t, \ldots, y_T)
= \min \set*{0, \min_{y_{t+1}, \ldots, y_{T-1} \in \R} \mathcal{C}_t(y_t, \ldots, y_{T-1}, \theta)}.
\end{align}
Define
\begin{align}
    C &\coloneqq \sum_{i=t+1}^T\delta^{-\frac{i-t}{\alpha -1}} = \delta^{-\frac{T-t+1}{\alpha-1}} \sumdelta{T-t},
    \quad u_i \coloneqq \frac{\delta^{-\frac{i-t}{\alpha -1}}}{C}.
\end{align}
Then, the second term of $\mathcal{C}_t$ can be bounded as follows:
\begin{align}
\sum_{i=t+1}^T \delta^{i-t} (y_i - y_{i-1})^{\alpha} &= \delta^{T-t+1} \sumdelta{T-t}^{-(\alpha-1)} \sum_{i=t+1}^T u_i^{-(\alpha-1)} (y_i - y_{i-1})^{\alpha} \\
&= \delta^{T-t+1} \sumdelta{T-t}^{-(\alpha-1)} \sum_{i=t+1}^T u_i \prn*{\frac{y_i - y_{i-1}}{u_i}}^{\alpha} \\
&\geq \delta^{T-t+1} \sumdelta{T-t}^{-(\alpha-1)} \prn*{\sum_{i=t+1}^T \prn*{y_i - y_{i-1}}}^{\alpha} \\
&= \delta^{T-t+1} \sumdelta{T-t}^{-(\alpha-1)} \prn*{\theta - y_t}^{\alpha},
\end{align}
where the inequality follows from Jensen's inequality. Equality holds if
\begin{align}
    \frac{y_i - y_{i-1}}{u_i} = \mathrm{constant}, \quad i = t+1, \ldots, T.
\end{align}
Using this, we obtain
\begin{align}
    c(y_{t} - x_{t-1}) + \sum_{i=t+1}^T \beta \delta^{i- t} c(y_i - y_{i-1}) 
    \ge (y_t - x_{t-1})^{\alpha} + \beta \delta^{T-t+1} \sumdelta{T-t}^{-(\alpha-1)} \prn*{\theta - y_t}^{\alpha}.
\end{align}
To further bound the right-hand side, we apply Hölder's inequality.

\begin{lemma}[Hölder's inequality] \label{lem:holder}
Let $p,q>0$ satisfy $\tfrac{1}{p} + \tfrac{1}{q} = 1$, and let $(a_i)_{i=1}^n$, $(b_i)_{i=1}^n$ be real sequences. Then
\begin{align}
    \prn*{\sum_{i=1}^n |a_i|^p}^{\frac{1}{p}}
    \prn*{\sum_{i=1}^n |b_i|^q}^{\frac{1}{q}}
    \ge \sum_{i=1}^n |a_i b_i|,
\end{align}
with equality if there exists a constant $D$ such that $|a_i|^p = D|b_i|^q$ for all $i$.
\end{lemma}

Setting $n=2$ and
\begin{align}
a_1 &= y_t - x_{t-1}, \quad 
a_2 = \beta^{\frac{1}{\alpha}} \delta^{\frac{T-t+1}{\alpha}} S_{T-t}^{-\frac{\alpha-1}{\alpha}} (\theta - y_t), \quad
b_1 = 1, \quad 
b_2 = \beta^{-\frac{1}{\alpha}} \delta^{-\frac{T-t+1}{\alpha}} S_{T-t}^{\frac{\alpha-1}{\alpha}}, \\
p &= \alpha, \quad q = \tfrac{\alpha}{\alpha-1},
\end{align}
we obtain
\begin{align}
    \prn*{(y_t - x_{t-1})^{\alpha} + \beta \delta^{T-t+1} \sumdelta{T-1}^{-(\alpha-1)} \prn*{\theta - y_t}^{\alpha}}^{\frac{1}{\alpha}}
    \prn*{1 + \beta^{-\frac{1}{\alpha-1}} \delta^{\frac{T-t+1}{\alpha -1}}  \sumdelta{T-t}}^{\frac{\alpha-1}{\alpha}}
    \ge \theta - x_{t-1},
\end{align}
hence
\begin{align}
    (y_t - x_{t-1})^{\alpha} + \beta \delta^{T-t+1} \sumdelta{T-1}^{-(\alpha-1)} \prn*{\theta - y_t}^{\alpha} \ge \prn*{1 + \beta^{-\frac{1}{\alpha-1}} \delta^{\frac{T-t+1}{\alpha -1}} \sumdelta{T-t}}^{-(\alpha-1)} \prn*{\theta - x_{t-1}}^{\alpha}.
\end{align}
Equality holds if and only if
\begin{align}
    (y_t - x_{t-1})^{\alpha} 
    \prn*{\beta^{-\frac{1}{\alpha}} \delta^{-\frac{T-t+1}{\alpha}} S_{T-t}^{\frac{\alpha-1}{\alpha}}}^{\frac{\alpha}{\alpha-1}} =
    \prn*{\beta^{\frac{1}{\alpha}} \delta^{\frac{T-t+1}{\alpha}} S_{T-t}^{-\frac{\alpha-1}{\alpha}} (\theta - y_t)}^{\alpha},
\end{align}
which is equivalent to
\begin{align}
y_t 
&= \frac{\sumdelta{T-t} x_{t-1} + \tbeta \tdelta^{T-t+1} \theta}{\sumdelta{T-t} + \tbeta \tdelta^{T-t+1}}
  \label{eq:y_t}. 
\end{align}

Consequently, when 
\begin{align}
    \prn*{1 + \beta^{-\frac{1}{\alpha-1}} \delta^{\frac{T-t+1}{\alpha -1}}  \sumdelta{T-t}}^{-(\alpha-1)} \prn*{\theta - x_{t-1}}^{\alpha}
    - \beta \delta^{T-t+1} R \le 0 \\
    \Leftrightarrow
    x_{t-1} 
    \ge \theta - R^{\frac{1}{\alpha}} \prn*{\sumdelta{T-t} + \tbeta \tdelta^{T-t+1}}^{\frac{\alpha-1}{\alpha}} \label{eq:case condition}
\end{align}
holds, the minimum in \cref{eq:min min} is attained by \cref{eq:y_t}, and otherwise the minimum is attained at $y_t = x_{t-1}$. Therefore, for $t=1,\ldots,T$,
\begin{align}
    x_t = 
    \begin{cases}
    \frac{\sumdelta{T-t} x_{t-1} + \tbeta \tdelta^{T-t+1} \theta}{\sumdelta{T-t} + \tbeta \tdelta^{T-t+1}}
    & \text{if \cref{eq:case condition} holds}, \\
    x_{t-1} & \text{otherwise}.
    \end{cases}
\end{align}

A simple calculation shows 
the right-hand side of \cref{eq:case condition}
is strictly increasing in $t$, so there exists $t^*\in\{0,\dots,T\}$ such that
\begin{align}
    x_t = 
    \begin{cases}
    \frac{\sumdelta{T-t} x_{t-1} + \tbeta \tdelta^{T-t+1} \theta}{\sumdelta{T-t} + \tbeta \tdelta^{T-t+1}} & t \le t^*, \\
    x_{t-1} & \text{otherwise},
    \end{cases}
\end{align}
and by unrolling this recursion we obtain
\begin{align}
x_t = \theta \prn*{1 - \prod_{i=1}^{\min\{t,t^*\}} p_i},
\end{align}
where $p_i = \tfrac{\sumdelta{T-i}}{\sumdelta{T-i} + \tbeta \tdelta^{T-i+1}}$. Finally, $t^*$ is the smallest $t\in\{0,\ldots,T-1\}$ satisfying
\begin{align}
\theta \prn*{1 - \prod_{i=1}^t p_i} 
< \theta - R^{\frac{1}{\alpha}} \prn*{\sumdelta{T-t-1} + \tbeta \tdelta^{T-t}}^{\frac{\alpha-1}{\alpha}},
\end{align}
which is easily seen to be equivalent to \cref{eq:condition_for_tast}.
\end{proof}

\section{Proof of \cref{thm:TAI condition}} \label{sec:proof of thm:TAI condition}

\begin{proof}
First, we prove the following lemma.
\begin{lemma}\label{lem:q_beta}
Fix $\alpha$, $\delta$, and $T$. Then the followings hold:
\begin{enuminlem}
  \item If $\beta \le h_1^{-1}(\delta)$, then 
    $
      q_0 < q_1 < \cdots < q_{T-1}.
    $
  \item If $h_1^{-1}(\delta) < \beta < h_2^{-1}(\delta)$, then there exists 
    $t \in \{0,\dots,T-1\}$ such that
    $
      q_0 \ge q_1 \ge \cdots \ge q_t
      < q_{t+1} < \cdots < q_{T-1}.
    $
  \item If $\beta \ge h_2^{-1}(\delta)$, then
    $
      q_0 > q_1 > \cdots > q_{T-1}.
    $
\end{enuminlem}
\end{lemma}

\begin{proof}
By the definition \eqref{eq:def qt} of $q_t$, we have
\begin{align}
\log \frac{q_t}{q_{t-1}} &= \log\prn*{\prn*{\frac{S_{T-t-1} + \tbeta\tdelta^{T-t}}
                           {S_{T-t}   + \tbeta\tdelta^{T-t+1}}}^{1-\alpha}
         \prn*{\frac{S_{T-t}}{S_{T-t} + \tbeta\tdelta^{T-t+1}}}^{\alpha}}\\
&= -(\alpha-1)\log\prn*{1 - \frac{1-\tbeta}{\frac{S_{T-t}}{\tdelta^{T-t}}}}
   -\log\prn*{1 + \frac{\tbeta\tdelta}{\frac{S_{T-t}}{\tdelta^{T-t}}}}.
\end{align}
Note that 
\begin{align}
  \frac{S_{T-t}}{\tdelta^{T-t}}
  = 1 + \frac1{\tdelta} + \frac1{\tdelta^2} + \cdots + \frac1{\tdelta^{T-t-1}}
\end{align}
decreases monotonically in $t$ and always lies in $[1,\infty)$ for $t=1,\dots,T-1$.
We investigate the function
\begin{align}
  f(x) \coloneqq -(\alpha-1)\log\prn*{1 - \tfrac{1-\tbeta}{x}}
              -\log\prn*{1 + \tfrac{\tbeta\tdelta}{x}}.
\end{align}
We have
\begin{align}
  &\lim_{x\searrow 1-\tbeta} f(x) = +\infty,\quad
  \lim_{x\to+\infty} f(x) = 0, \\
  &f'(x)
  = \frac{\prn*{\tbeta\tdelta - (\alpha-1)(1-\tbeta)}x
          - \alpha\tbeta\tdelta(1-\tbeta)}
         {x(x - (1-\tbeta))(x + \tbeta\tdelta)}
\end{align}
hold. 

\paragraph{Case (a): $\beta \le h_1^{-1}(\delta)$.}
In this case, $\tbeta\tdelta - (\alpha-1)(1-\tbeta)\le0$ and $\alpha\tbeta\tdelta(1-\tbeta)\ge0$, so $f'(x)\le0$ for all $x>1-\tbeta$. Because $f$ is decreasing on $(1-\tbeta,\infty)$, we get $f(x)>0$ for all $x\ge1$. Therefore $\log(q_t/q_{t-1})>0$ for $t=1,\dots,T-1$, proving (a).

\paragraph{Case (b): $h_1^{-1}(\delta) < \beta < h_2^{-1}(\delta)$.}
In this case, $\tbeta\tdelta - (\alpha-1)(1-\tbeta)>0$, and we can check $f'(1)<0$. 
Thus $f'(x)\le0$ for $x \leq \frac{\alpha\tbeta\tdelta(1-\tbeta)}{\tbeta\tdelta - (\alpha-1)(1-\tbeta)}$ and $f'(x)>0$ for $x> \frac{\alpha\tbeta\tdelta(1-\tbeta)}{\tbeta\tdelta - (\alpha-1)(1-\tbeta)}$.
So, there is a unique threshold $\tilde x < 1 - \tbeta$ and $f(x)>0$ for $x<\tilde x$ and $f(x)\le0$ for $x\ge\tilde x$. Since $S_{T-t}/\tdelta^{T-t}$ decreases in $t$, this yields the claimed pattern of monotonicity in (b).

\paragraph{Case (c): $\beta \ge h_2^{-1}(\delta)$.}
In this case $\tbeta\tdelta - (\alpha-1)(1-\tbeta)\ge0$ and $f'(1)\ge0$, so $f'(x)\ge0$ for all $x\ge1$. Hence $f$ is strictly increasing on $[1,\infty)$, we have $f(x)<0$ for $x\ge1$. Therefore $\log(q_t/q_{t-1})<0$ for all $t=1,\dots,T-1$, proving (c).
\end{proof}

Next, through a more detailed analysis of the case $h_1^{-1}(\delta) < \beta < h_2^{-1}(\delta)$, we obtain the following lemma. 

\begin{lemma}\label{lem:q_beta_closer}
Fix $\alpha$, $\delta$, and $T$. There exists a unique
\begin{align}
  \beta_0 \in \prn*{h_1^{-1}(\delta),h_2^{-1}(\delta)}
\end{align}
such that:
\begin{enumerate}
  \item If $h_1^{-1}(\delta) < \beta < \beta_0$, then $q_0 < q_{T-1}$.
  \item If $\beta = \beta_0$, then $q_0 = q_{T-1}$.
  \item If $\beta_0 < \beta < h_2^{-1}(\delta)$, then $q_0 > q_{T-1}$.
\end{enumerate}
\end{lemma}

\begin{proof}
Observe that
\begin{align}
  \frac{q_{T-1}}{q_0}
  = \prn*{\tfrac{S_{T-1}}{\tbeta\tdelta} + \tdelta^{T-1}}^{\alpha-1}
    \prod_{t=1}^{T-1}\prn*{\tfrac{S_{T-t}}{S_{T-t} + \tbeta\tdelta^{T-t+1}}}^{\alpha}
\end{align}
is strictly decreasing in $\beta$. By \cref{lem:q_beta}, at $\beta = h_1^{-1}(\delta)$ we have $q_{T-1}/q_0>1$, and at $\beta = h_2^{-1}(\delta)$ we have $q_{T-1}/q_0<1$. The intermediate value theorem yields a unique $\beta_0$ satisfying $q_{T-1}/q_0=1$, and the proof completes. 
\end{proof}

Finally, to prove \cref{thm:TAI condition}, recall that $(\beta,\delta)$ is TAI if and only if 
\begin{align}
  \max_{t=0,\dots,T-1} q_t \neq q_0
\end{align}
by \eqref{eq:TAI q_t condition}. Combining \cref{lem:q_beta} and \cref{lem:q_beta_closer} completes the proof.
\end{proof}

\section{Proof of \cref{thm:optimal_goal_non_exploitative}}
Since exploitative rewards are disallowed, the agent must never abandon the task, which requires
$
  \max_{t\in\{0,\dots,T-1\}} q_t \;\le\; \frac{R}{\theta^\alpha}.
$
So, setting 
\begin{align} \label{eq:optimal theta}
\theta = \prn*{\frac{R}{\max_{t\in\{0,\dots,T-1\}} q_t}}^{\frac{1}{\alpha}}
\end{align}
is optimal. 
If $(\beta,\delta)$ is TAI, then $\max_{t}q_t = q_{T-1}$, and if it is not TAI, then $\max_{t}q_t = q_0$. Substituting these cases into \cref{eq:optimal theta} yields \cref{eq:optimal_goal_non_TAI} and \cref{eq:optimal_goal_TAI}, respectively.

\section{Proof of \cref{thm:optimal_goal_exploitative}}
In the non-TAI case, the final progress satisfies
\begin{align}
  x_T = 
  \begin{cases}
    \theta , & \text{if } q_0 \le \dfrac{R}{\theta^{\alpha}},\\
    0,           & \text{otherwise}.
  \end{cases}
\end{align}
Therefore, maximizing $x_T$ reduces to choosing $\theta$ as large as possible under the constraint $q_0 \le \tfrac{R}{\theta^{\alpha}}$, which yields the optimal solution
$
  \theta = \prn*{\frac{R}{q_0}}^{\frac{1}{\alpha}}.
$

Next, we consider the TAI case. Let $t^*$ denote the time at which the agent abandons the task (see Theorem~\cref{thm:state sequence}). We distinguish three cases:

\paragraph{Case (a): $t^* = 0$.}
In this case, the final progress satisfies
\begin{align}
  x_T = 0.
\end{align}

\paragraph{Case (b): $t^* = T$.}
Here the agent never abandons the task, so the outcome coincides with the non-exploitative case. Hence
\begin{align}
  \theta &= \prn*{\frac{R}{q_{T-1}}}^{\frac{1}{\alpha}},\\
  x_T    &= \theta.
\end{align}

\paragraph{Case (c): $t^* \in \{1,\ldots,T-1\}$.}
The necessary and sufficient conditions for abandonment at time $t^*$ are
\begin{align}
  q_s      &\le \frac{R}{\theta^\alpha} \quad s = 0,\ldots,t^*-1, \label{eq:tstar1}\\
  q_{t^*}  &>  \frac{R}{\theta^\alpha}.            \label{eq:tstar2}
\end{align}
It follows that
\begin{align}
  q_{t^*} > \frac{R}{\theta^\alpha} \ge q_0,
\end{align}
where the first inequality uses \cref{eq:tstar2} and the second uses \cref{eq:tstar1}. Since $q_{t^*}>q_0$, if
\begin{align}
  \prn*{\frac{R}{q_{t^*}}}^{\frac{1}{\alpha}}
  < \theta
  \le
  \prn*{\frac{R}{\max_{s=0,\ldots,t^*-1}q_s}}^{\frac{1}{\alpha}} = \prn*{\frac{R}{\max\{q_0,q_{t^*-1}\}}}^{\frac{1}{\alpha}},
\end{align}
the agent abandons at $t^*$. The resulting final progress is
\begin{align}
  x_T = \theta \,\prn*{1 - \prod_{i=1}^{t^*}p_i}.
\end{align}
Because this is increasing in $\theta$, the optimal choice is
\begin{align}
  \theta
  = \prn*{\frac{R}{\max\{q_0,q_{t^*-1}\}}}^{\frac{1}{\alpha}},
\end{align}
and thus
\begin{align}
  x_T
  = \prn*{\frac{R}{\max\{q_0,q_{t^*-1}\}}}^{\frac{1}{\alpha}}
    \prn*{1 - \prod_{i=1}^{t^*}p_i}.
\end{align}

Combining Cases (a)--(c), the goal-setting problem reduces to
\begin{align}
  \max_{t \in \{\tilde t,\ldots,T\}}
  \prn*{\frac{R}{\max\{q_0,q_{t-1}\}}}^{\frac{1}{\alpha}}
  \prn*{1 - \prod_{i=1}^t p_i},
\end{align}
where $\tilde t$ is the smallest $t$ satisfying $q_t>q_0$. 
Using $t$T that achieves the maximum, the optimal goal is then
\begin{align}
  \theta = \prn*{\frac{R}{\max\{q_0,q_{t-1}\}}}^{\frac{1}{\alpha}}. 
\end{align}

\section{Proof of \cref{thm:optimal_reward_scheduling_non_TAI}}
\begin{proof}
Because exploitative rewards are not allowed, when we fix $T_i$ and $R_i$, \cref{thm:optimal_goal_non_exploitative} implies that the optimal choice of $\theta_i$ is
\begin{align}
  \theta_i 
  &= R_i^{\frac{1}{\alpha}} \prn*{F(T_i)}^{\frac{\alpha-1}{\alpha}}. 
\end{align}
Substituting into the objective gives
\begin{align}
  \sum_{i=1}^N \theta_i
  &= \sum_{i=1}^N R_i^{\frac{1}{\alpha}} \prn*{F(T_i)}^{\frac{\alpha-1}{\alpha}} \le \prn*{\sum_{i=1}^N R_i}^{\frac{1}{\alpha}} \prn*{\sum_{i=1}^N F(T_i)}^{\frac{\alpha-1}{\alpha}} = R^{\frac{1}{\alpha}} \prn*{\sum_{i=1}^N F(T_i)}^{\frac{\alpha-1}{\alpha}},
\end{align}
where the inequality follows by Hölder's inequality (\cref{lem:holder}). Equality holds if and only if
\begin{align}
  R_i \propto F(T_i). 
\end{align}
for all $i$.

Therefore, it is sufficient to maximize $\sum_{i=1}^N F(T_i)$ under constraints, and the optimal reward scheduling problem reduces to problem \cref{eq:problem_reward_scheduling_reduced}.
\end{proof}

\section{Properties of $h_1^{-1}$ and $h_2^{-1}$} \label{sec:properties of h_inv}
We define 
\begin{align}
g^-(\delta) &\coloneqq \sqrt{(\alpha - 1)^2\prn*{1 - \delta^{\frac{1}{\alpha - 1}}}^2
      + 4\alpha(\alpha - 1)\delta^{\frac{1}{\alpha - 1}}} \\
      &-(\alpha - 1)\prn*{1 - \delta^{\frac{1}{\alpha - 1}}}, \\
g^+(\delta) &\coloneqq \sqrt{(\alpha - 1)^2\prn*{1 - \delta^{\frac{1}{\alpha - 1}}}^2
      + 4\alpha(\alpha - 1)\delta^{\frac{1}{\alpha - 1}}} \\
      &+(\alpha - 1)\prn*{1 - \delta^{\frac{1}{\alpha - 1}}}. 
\end{align}

\subsection{Closed Form Expression.}
The functions $h_1^{-1}$ and $h_2^{-1}$ can be expressed explicitly as functions of $\delta$ as follows:
\begin{align}
  h_1^{-1}(\delta)
    &= \prn*{\frac{\alpha - 1}{\alpha - 1 + \delta^{\frac{1}{\alpha - 1}}}}^{\alpha - 1},\\
  h_2^{-1}(\delta)
    &= \prn*{\frac{g^-(\delta)}
      {2\alpha\delta^{\frac{1}{\alpha - 1}}}}^{\alpha - 1}
    = \prn*{\frac{2 (\alpha - 1)}
      {g^+(\delta)}}^{\alpha - 1} .
\end{align}

\subsection{Limit as $\alpha \to 1^+$. }
When $0 < \delta < 1$, because $\lim_{\alpha \to 1^+} \delta^{\frac{1}{\alpha - 1}} = 0$, 
\begin{align}
  \lim_{\alpha \to 1^+} h_1^{-1}(\delta) &= 1,  \\
  \lim_{\alpha \to 1^+} h_2^{-1}(\delta) &= 1
  . 
\end{align}
When $\delta = 1$,  because $\lim_{\alpha \to 1^+} \delta^{\frac{1}{\alpha - 1}} = 1$, 
\begin{align}
  \lim_{\alpha \to 1^+} h_1^{-1}(\delta)
  &= \lim_{\alpha \to 1^+} \frac{(\alpha-1)^{\alpha-1}}{\alpha^{\alpha-1}} = 1,  \\
  \lim_{\alpha \to 1^+} h_2^{-1}(\delta) &= 
  \lim_{\alpha \to 1^+} \sqrt{\frac{(\alpha-1)^{\alpha-1}}{\alpha^{\alpha-1}}} = 1
  . 
\end{align}
From the above, we have 
\begin{align}
  \lim_{\alpha \to 1^+} h_1^{-1}(\delta) = \lim_{\alpha \to 1^+} h_2^{-1}(\delta) &= 1
\end{align}
for $0 < \delta \leq 1$. 

\subsection{Limit as $\alpha \to \infty$. }
Because $\lim_{\alpha \to \infty} \delta^{\frac{1}{\alpha - 1}} = 1$ for $0 < \delta \leq 1$, 
\begin{align}
  \lim_{\alpha \to \infty} h_1^{-1}(\delta) &= \lim_{\alpha \to \infty}  \prn*{\frac{\alpha-1}{\alpha}}^{\alpha - 1} = \frac{1}{e}, \\
  \lim_{\alpha \to \infty} h_2^{-1}(\delta) &= \lim_{\alpha \to \infty}  \sqrt{\prn*{\frac{\alpha-1}{\alpha}}^{\alpha - 1}} = \frac{1}{\sqrt{e}}. 
\end{align}

\section{Dynamic Programming Algorithm for \cref{eq:problem_reward_scheduling_reduced}} \label{sec:reward-scheduling-dp}
\renewcommand{\algorithmicrequire}{\textbf{Input:}}
\renewcommand{\algorithmicensure}{\textbf{Output:}}

The dynamic programming algorithm for \cref{eq:problem_reward_scheduling_reduced} is described in \cref{alg:reward-scheduling-dp}. 

\begin{algorithm}[h]
\caption{Dynamic Programming for Problem \cref{eq:problem_reward_scheduling_reduced}}
\label{alg:reward-scheduling-dp}
\begin{algorithmic}[1]
\REQUIRE Total period $T$, function $F$
\ENSURE Optimal segments $(T_1,\dots,T_k)$
\STATE Initialize $v[0]\gets 0$ and $v[t]\gets -\infty$ for $t=1,\dots,T$
\FOR{$t=1$ to $T$}
  \FOR{$s=1$ to $t$}
    \IF{$F(s)+v[t-s]>v[t]$}
      \STATE $v[t]\gets F(s)+v[t-s]$
      \STATE $\mathit{prev}[t]\gets s$
    \ENDIF
  \ENDFOR
\ENDFOR
\STATE $t\gets T$
\STATE $\mathit{segments}\gets []$
\WHILE{$t>0$}
  \STATE $s\gets \mathit{prev}[t]$
  \STATE \textbf{prepend} $s$ to $\mathit{segments}$
  \STATE $t\gets t-s$
\ENDWHILE
\RETURN $\mathit{segments}$
\end{algorithmic}
\end{algorithm}

\section{Details of Numerical Experiments}
All the numerical experiments were conducted on a MacBook Pro (14-inch, 2023) with Apple M2 Max, 32 GB memory, running macOS Sonoma 14.5. 
All the codes are implemented in Python 3.11.2. 
Because all numerical experiments are deterministic, we report the results of a single run.

\end{document}